\documentclass[11pt,onecolumn,amssymb,nofootinbib]{revtex4}
\usepackage{amsmath, amsthm, amscd, amssymb}
\usepackage{graphicx, braket}
\usepackage{bm}
\usepackage{bbm}

\let\vec\boldvec%
\newcommand*{\ii}{\ensuremath{\mathrm{i}}}

\DeclareMathOperator{\linearspan}{span}

\begin{document}

\title{\bf The Hilbert space operator formalism within dynamical reduction
models}
\author{Angelo Bassi}
\email{bassi@ts.infn.it}
\address{Dipartimento di Fisica Teorica,
Universit\`a di Trieste, Strada Costiera 11, 34014 Trieste, Italy.
\\  Mathematisches Institut der L.M.U., Theresienstr. 39, 80333
M\"unchen, Germany, \\
Istituto Nazionale di Fisica Nucleare, Sezione di Trieste, Strada
Costiera 11, 34014 Trieste, Italy.}

\author{GianCarlo Ghirardi}
\email{ghirardi@ts.infn.it}%
\affiliation{Dipartimento di Fisica Teorica dell'Universit\`a degli
Studi di Trieste, Strada Costiera 11, 34014 Trieste, Italy, \\
Istituto Nazionale di Fisica Nucleare, Sezione di Trieste, Strada
Costiera 11, 34014 Trieste, Italy, \\ The Abdus Salam International
Centre for Theoretical Physics, Strada Costiera 11, 34014 Trieste,
Italy}

\author{Davide~G.~M. Salvetti}
\email{salvetti@ts.infn.it} \affiliation{Dipartimento di Fisica
Teorica dell'Universit\`a degli Studi di Trieste, Strada Costiera
11, 34014 Trieste, Italy, \\ Istituto Nazionale di Fisica Nucleare,
Sezione di Trieste, Strada Costiera 11, 34014 Trieste, Italy}

\begin{abstract}
Unlike standard quantum mechanics, dynamical reduction models assign
no particular \emph{a priori} status to ``measurement processes'',
``apparata'', and ``observables'', nor self-adjoint operators and
positive operator valued measures enter the postulates defining
these models. In this paper, we show why and how the Hilbert-space
operator formalism, which standard quantum mechanics
\emph{postulates}, can be \emph{derived} from the fundamental
evolution equation of dynamical reduction models. Far from having
any special ontological meaning, we show that within the dynamical
reduction context the operator formalism is just a compact and
convenient way to express the statistical properties of the outcomes
of experiments.
\end{abstract}
\maketitle

\section{Introduction}
\label{sec:introduction}

Dynamical Reduction Models~(DRMs) provide, at the non relativistic
level at least, a coherent unified description of both microscopic
quantum and macroscopic classical phenomena, and in particular give
a consistent solution to the macro-objectification problem of
quantum mechanics~\cite{grw,rev1}. They are defined by the
following set of axioms:\\

\noindent {\textsc{Axiom A: states.} A Hilbert space ${\mathcal H}$
is associated to any physical system and the state of the system is
represented by a (normalized) vector $|\psi\rangle$ in ${\mathcal
H}$.\\

\noindent \textsc{Axiom B: evolution (GRW model).} At random times,
distributed like a Poissonian process with mean frequency $\lambda$,
each particle of a system of $N$ particles is subjected to a
spontaneous localization process of the form:
\begin{equation} \label{eq:sdfsd}
|\psi_{t}\rangle \; \longrightarrow \; \frac{L_{n}({\bf x})
|\psi_{t}\rangle}{\| L_{n}({\bf x}) |\psi_{t}\rangle \|},
\qquad\qquad L_{n}({\bf x}) \; = \;
\sqrt[4]{\left(\frac{\alpha}{\pi}\right)^3}\, \exp\left[ -
\frac{\alpha}{2} ({\bf q}_{n} - {\bf x})^2 \right],
\end{equation}
where ${\bf q}_{n}$ is the position operator associated to the
$n$-th particle, and $|\psi_{t}\rangle$ is the wave function of the
global system immediately prior to the collapse; the collapse
processes for different particles are independent. Between two
collapses, the wave function evolves according to the standard
Schr\"odinger equation. The probability density for a collapse for
the $n$-th particle to occur around the point ${\bf x}$ of space is:
\begin{equation} \label{eq:ghhfd}
p({\bf x}) \quad = \quad \| L_{n}({\bf x}) |\psi_{t}\rangle \|^2.
\end{equation}
The standard numerical values \cite{grw} for the two parameters
$\lambda$ and $\alpha$ are: $\lambda \simeq 10^{-16}$ sec$^{-1}$ and
$\alpha \simeq 10^{10}$ cm$^{-2}$. A continuous formulation in terms
of stochastic differential equations is also commonly
used~\cite{rev1,rev2,pp,gpr,rev3}.\\

\noindent {\textsc{Axiom C: ontology.} Let $\psi({\bf x}_{1}, {\bf
x}_{2}, \ldots {\bf x}_{N}) \equiv \langle {\bf x}_{1}, {\bf x}_{2},
\ldots {\bf x}_{N}| \psi\rangle$ the wave function for a system of
$N$ particles (which for simplicity we take to be scalar) in
configuration space. Then
\begin{equation}
m({\bf x},t) \; \equiv \; \sum_{n=1}^{N} m_{n} \int d^3 x_{1} \ldots
d^3 x_{N} \, \delta^{(3)}({\bf x}_{n} - {\bf x})| \psi({\bf x}_{1},
{\bf x}_{2}, \ldots {\bf x}_{N}) |^2
\end{equation}
is assumed to describe the {\it density of mass}\footnote{In the
subsequent sections, for simplicity's sake, we will not make
reference to the mass density function anymore, and we will only
keep track of the evolution of the wave function; however it should
be clear that, in order to be fully rigorous, all statements about
the properties of physical systems should be phrased in terms of
their mass-density distribution.} distribution of the
system\footnote{The mass density in principle refers to the whole
universe; however, as standard practice in Physics, one can make the
approximation of considering only a part of it, which is
sufficiently well isolated, and of ignoring the state of the rest of
the universe.} in three dimensional space, as a function of
time~\cite{rev1,ggb}.

As we see, within DRMs concepts like {\it measurement}, {\it
apparata}, {\it observables} play no particular privileged role;
like in classical mechanics, they merely refer to particular
physical situations where a macroscopic physical system, which we
call apparatus, interacts in a specific way with another physical
system, e.g. a microscopic quantum system. However, such a
macroscopic system, the apparatus, is ultimately described in terms
of its fundamental constituents, and its interaction with other
systems is ultimately described in terms of the fundamental
interactions of nature. The basic idea should be clear: all physical
processes are governed by the universal dynamics embodied in the
precise axioms we have just presented. What is usually denoted as a
"measurement" of an observable by an apparatus is simply a precise
physical process which is purposely caused by a human being under
controlled conditions. In what follows we will use, to denote such a
situation, the term "experiment" to conform to the clear-cut
position of J.S. Bell, summarized in the following lucid
sentence~\cite{bell}:
\begin{quote}
I am convinced that the word `measurement' has now been so abused
that the field would be significantly advanced by banning its use
altogether, in favor, for example, of the word `experiment'.
\end{quote}

Given this, the following question arises: why are experiments on
microscopic quantum systems so efficiently described in terms of
average values of self-adjoint operators, and more generally in
terms of POVMs? Why is the Hilbert space formalism so powerful in
accounting for the observable properties of microscopic systems? The
aim of this paper is to provide an answer to these questions, from
the point of view of DRMs. We will show that, within DRMs, one can
{\it derive} a well defined role for self-adjoint operators and
POVMs as useful (but not compelling) mathematical tools which allow
to compactly express the statistical properties of microscopic
systems subject to experiments. Accordingly, within DRMs, recovering
the formal aspects of standard quantum mechanics is simply a matter
of practical convenience and mathematical elegance. Stated in
different terms, while experiments on quantum systems are nothing
more than a particular type of interaction between a macroscopic
(thus always well localized) system and a microscopic one, such that
different macroscopic configurations of the macro-object correspond
to different outcomes of the experiment, it is nevertheless simpler
to refer to the statistical properties of the outcomes in terms of
self-adjoint operators averaged over the initial state of the
microscopic quantum system. Just a matter of practical convenience,
nothing more.

The paper is organized as follows.  In Sec.~\ref{sec:hsof:example}
we show with a simple example how DRMs recover the operator
formalism for the description of the statistical properties of the
outcomes of quantum experiments. Sec.~\ref{sec:hsof-drm} is the core
of the paper: we will prove in full generality how DRMs allow to
associate a POVM to an experiment, i.e. how the statistical
properties of the experiment can be represented as the average
values of the effects of the POVM over the state before the
``measurement" of the microscopic quantum system. In
Sec.~\ref{sec:hsof:reproducibility} we will show that when an
experiment is reproducible, the POVM reduces to a PVM and the
experiment can be represented by a unique self-adjoint operator, as
it is typically assumed in standard text books on quantum mechanics.
In Sec.~\ref{sec:hsop-drm:stern_gerlach_experiment} we reconsider
the so-called ``tail'' problem and show that it does not affect the
dynamical reduction program. In Sec.~\ref{sec:cla} we show with an
explicit example that the Hilbert space formalism can be used also
to describe certain {\it classical} experiments.
Sec.~\ref{sec:conclusions} contains some concluding remarks.

This paper takes most inspiration from Ref.~\cite{bm}, where the
emergence of the operator formalism within the framework of Bohmian
Mechanics has been thoroughly analyzed.

\section{Emergence of the operator formalism: a simple example}
\label{sec:hsof:example}

We begin our analysis by discussing a simple physical situation
which should make clear how, and in which sense, the operator
formalism of standard quantum mechanics ``emerges'' in a natural way
from the physical and mathematical properties of DRMs, even if it
does not appear explicitly in the axioms defining these models.

\subsection{A measurement situation}

For simplicity's sake, let us consider a spin-1 particle (its
Hilbert space being~${\mathcal H} = {\mathbb C}^3$) and let us
assume that it has been initially prepared in the normalized state
\begin{equation} \label{eq:97}
\Ket{\psi} = a\Ket{S_{z}=+1} + b\Ket{S_{z}=0} + c\Ket{S_{z}=-1},
\end{equation}
where $\ket{S_{z}=+1}$, $\ket{S_{z}=0}$, and~$\ket{S_{z}=-1}$ are
the three eigenstates\footnote{There is of course nothing special in
the choice of the eigenstates of~$S_{z}$: we could have very well
chosen another basis of~${\mathbb C}^3$.} of~$S_{z}$, and $a$, $b$,
and~$c$ are three complex parameters which can be varied according
to the preparation procedure. Let us now perform a Stern-Gerlach
type of experiment which measures the spin of the particle along the
$x$ direction. According to the rules of
DRMs\footnote{See~\cite{mis} for a details analysis of this topic.}
(no other assumption is used, other than axioms A--C, in the
simplified form suitable for this example), we can state that:
\begin{itemize}
\item Throughout the entire process, the measuring device has
always a well defined macroscopic configuration. In the particular
example we intend to discuss, there are three possible outcomes for
the experiment, i.e. three possible final macroscopically different
configurations of the measuring device, which we call e.g. ``outcome
$+1$'', ``outcome 0'' and ``outcome $-1$'';

\item The outcome is random and the probability
distribution depends only on the initial state of the micro-system,
according to a law which, with great accuracy, is equal to:
\begin{eqnarray}
  \label{eq:104}
  {\mathbb P}^{\psi}(+1) & = &
  \frac{1}{4}|a + \sqrt{2} b + c|^{2},\\
  \label{eq:105}
  {\mathbb P}^{\psi}(0) & = &
  \frac{1}{2}|a - c|^{2},\\
  \label{eq:106}
  {\mathbb P}^{\psi}(-1) & = &
  \frac{1}{4}|a - \sqrt{2} b + c|^{2}.
\end{eqnarray}
\end{itemize}
It is worthwhile stressing that the above probabilities, which
ultimately coincide with the standard quantum probabilities
associated to the outcomes of a measurement of the spin along the
$x$ direction, are {\it not} postulated, but derive from the
dynamics of DRMs, when applied to the specific measurement-like
situation which has been chosen. Given these premises, we now show
how one can associate an operator, namely the spin operator $S_{x}$,
to this specific experiment.

\subsection{Observables as operators}

Let ${\mathcal O} = \set{-1, 0, +1}$ be the set of the possible
outcomes of the experiment, and let ${\mathcal G}$ be the power set
of ${\mathcal O}$ (${\mathcal G} = {\mathcal P}({\mathcal O})$),
which is an algebra on ${\mathcal O}$. We can then define a
probability measure on ${\mathcal G}$ in the following obvious way:
\begin{equation} \label{eq:probm}
{\mathbb P}^{\psi}(V) \;\; \equiv \;\; \sum_{v \in V} {\mathbb
P}^{\psi}(v) \qquad \forall \, V  \in  {\mathcal G},
\end{equation}
with ${\mathbb P}^{\psi}(v)$ given by
Eqs.~\eqref{eq:104}--\eqref{eq:106}. The above definition assumes
that $|\psi\rangle$ is a fixed unit vector, while $V$ can vary. Let
us now reverse the roles of $|\psi\rangle$ and $V$: we fix an
element $V \in {\mathcal G}$ and consider ${\mathbb P}^{\psi}(V)$ as
a function of $|\psi\rangle$. It is easy to recognize that the
probability distribution depends {\it quadratically} on the initial
state $|\psi\rangle$ (i.e. on the coefficients $a, b, c$) of the
micro-system: as a matter of fact, this is the very reason why one
can attach an operator to the experiment. Such a property is even
more evident if we introduce the following normalized and orthogonal
states:
\begin{eqnarray}
  \label{eq:98}
  \Ket{v=+1} &=& \frac{1}{2} \left[\Ket{S_{z}=+1}
    + \sqrt{2}\Ket{S_{z}=0} + \Ket{S_{z}=-1}\right],\\
  \label{eq:99}
  \Ket{v=0}  &=& \frac{1}{\sqrt{2}} \left[\Ket{S_{z}=+1} -
  \Ket{S_{z}=-1} \right],\\
  \label{eq:100}
  \Ket{v=-1} &=& \frac{1}{2} \left[\Ket{S_{z}=+1}
    - \sqrt{2}\Ket{S_{z}=0} + \Ket{S_{z}=-1} \right],
\end{eqnarray}
resorting to which we can express~\eqref{eq:probm} in the following
compact way:
\begin{equation} \label{eq:bvnv}
{\mathbb P}^{\psi}(V) \; = \; \sum_{v \in V} |\langle v | \psi
\rangle|^2.
\end{equation}
If we allow $|\psi\rangle$ to run over the entire Hilbert space
${\mathcal H}$, not just over the unit sphere\footnote{Of course,
${\mathbb P}^{\psi}(V)$ can be consistently interpreted as a
probability only when $|\psi\rangle \in {\mathcal S}^{1}$.}
${\mathcal S}^{1}$, then ${\mathbb P}^{\psi}(V)$ becomes a {\it
quadratic} function from ${\mathcal H}$ to ${\mathbb R}$, being the
diagonal part of the sesquilinear form
\begin{equation}
{\mathbb P}^{\psi, \phi}(V) \; = \; \sum_{v \in V}
\langle\psi|v\rangle \langle v|\phi\rangle \qquad \psi,\phi \, \in
\, {\mathcal H}, \quad V \; \text{fixed}.
\end{equation}
Given the bounded sesquilinear form ${\mathbb P}^{\psi, \phi}(V)$,
the Riesz representation theorem  allows us to express it in terms
of a bounded linear operator $O_{V}$, in the following way:
\begin{equation}
{\mathbb P}^{\psi, \phi}(V) \; = \; \langle \psi| O_{V} |\phi
\rangle, \qquad\quad ( \text{obviously, in this simple case:} \;\;
O_{V} \; \equiv \; \sum_{v \in V} |v\rangle\langle v| ).
\end{equation}
In this particular example we know also that the operator $O_{V}$ is
self-adjoint.

Going back to the original ${\mathbb P}^{\psi}$, we can then write:
\begin{equation}
{\mathbb P}^{\psi}(V) \; = \;  \langle \psi| O_{V} | \psi \rangle,
\qquad\quad \forall \,\, V \in {\mathcal G}, \; \forall \,\,
|\psi\rangle \in {\mathcal H},
\end{equation}
and the set of operators $\set{O_{V}}_{V \in {\mathcal G}}$ forms a
POVM, as one can easily prove. This is the result we wanted to
arrive at: thanks to the particular dependence of the probability
measure ${\mathbb P}^{\psi}$ on $\psi$ and to the Riesz
representation theorem we have been able to express the statistical
properties of the outcomes of the experiment in terms of average
values of the effects of a POVM over the initial state of the
microscopic system.

For our particular experiment, the set $\set{O_{V}}_{V \in {\mathcal
G}}$ is more than a POVM: each of the eight operators $O_{V}$ is in
fact a projection operator, and therefore $\set{O_{V}}_{V \in
{\mathcal G}}$ turns out to be a Projection Valued Measure~(PVM),
which is the one associated to the spectral resolution of a
self-adjoint operator
\begin{equation} \label{eq:118}
O \; = \; \sum_{v \in {\mathcal O}} v\, |v\rangle\langle v| \quad
\Rightarrow \quad O \; = \; \frac{1}{\sqrt{2}} \left(
\begin{array}{ccc}
0 & 1 & 0\\ 1 & 0 & 1\\ 0 & 1 & 0
\end{array}
\right).
\end{equation}
For this reason, we can rightly associate the operator~$O$ to our
experiment, in the precise sense given here. By inspecting the
components of the three vectors $\ket{v}$ given by
Eqs.~\eqref{eq:98}, \eqref{eq:99}, and \eqref{eq:100} and the
explicit form of the operator~$O$ displayed in Eq.~\eqref{eq:118},
we can now finally recognize that~$O$ is indeed the
component~$S_{x}$ of the spin along the ${x}$ direction for a spin 1
particle, written in the basis ${\ket{S_{z}=+1}, \ket{S_{z}=0},
\ket{S_{z}=-1}}$ of the eigenstates of~$S_{z}$.

\section{The Operator Formalism within Dynamical Reduction Models}
\label{sec:hsof-drm}

In this section, we prove in full generality what we have shown with
the previous example, namely how the Hilbert-space operator
formalism derives from DRMs as a tool to express the statistical
properties of the outcomes of the experiments.

\subsection{The link between experimental outcomes and macroscopic
positions} \label{sec:hsop-drm:outcomes_macrostates-link}

When describing physical experiments, one usually identifies
experimental outcomes with \emph{real numbers}; however, what one
actually sees (his empirical experience) as the outcome of a
measurement is not a real number, but a specific configuration of a
macroscopic object, namely a pointer being located in a well-defined
region of space\footnote{One might also consider different
situations, like e.g. the firing of a counter; however, what matters
is that in all measurement situations the final states of the
apparatus differs for a macroscopic mass density distribution. The
reader will have no difficulty in transcribing the following
analysis in such a way that it applies also to the just mentioned
situations in which there is not any pointer.}. It is then necessary
to set a link between real numbers interpreted as the outcome of an
experiment and the position of the pointer of the measuring
apparatus. We now wish to make precise the conditions which the
state of the pointer satisfies whenever we experience a
\emph{perception} which we interpret as: ``The outcome of the
experiment is~$v$'', $v$ being a real number.

\subsubsection{The position of a macroscopic object}
\label{sec:hsop-drm:macroscopic_objects-positions}

For simplicity's sake, we will analyze the pointer by considering
only the spatial degrees of freedom of its center of mass, ignoring
its spatial extension and orientation, as well as all its
microscopic degrees of freedom. According to the ontology of DRMs, a
pointer is a distribution of mass which---being macroscopic---is
localized within a small region of space coinciding with the spatial
extension of the pointer itself, but nevertheless it has small
``tails'' spreading out to infinity; because of this, it is not
possible to adopt the point of view according to which the pointer
(and, in general, any macroscopic object) is located in some given
region of space if and only if its mass density is entirely
contained within that region; we will come back to this point in
Sec.~\ref{sec:hsop-drm:stern_gerlach_experiment}. We then say that a
macro-object is located within a given region of space when almost
all its mass density distribution is contained within that region.

As a consequence of the dynamical laws of DRMs, in all standard
physical situations\footnote{Here we do not take into account those
pathological situations (whose probability of occurring is
vanishingly small) in which a macroscopic object can be, for a very
short time, in a superpositions of macroscopically different
states.} such as measurement processes, the region around which the
wave function of the center of mass of a macro-object is localized
is extremely small; accordingly, one can also take as the position
of a macro-object the center of that region, which can be
mathematically expressed by the formula
\begin{equation} \label{eq:283}
{\bf q}_{t} = \langle \psi_{t} | {\bf q} | \psi_{t}\rangle \; \in \;
{\mathbb R}^3,
\end{equation}
where~${\bf q}$ is the position operator of the center of mass
of~$A$. We stress that~${\bf q}_{t}$ should {\it not} be interpreted
as the quantum average of the position operator~${\bf q}$, as done
in standard quantum mechanics, but as the coordinates of the point
in space around which, at time $t$, the mass density is appreciably
different from zero. In the following, we will use~\eqref{eq:283} to
denote the position of a macro-object.

Coming back to our pointer, let us assume that the pointer moves
only along the graduated scale, so that we can treat it as a one
dimensional system: we will call $q_{t} \in {\mathbb R}$ its
position along the scale. Since its dynamical evolution is
intrinsically stochastic,~$q_{t}$ is a random variable~$q_{t}:
\Omega \rightarrow {\mathbb R}$ for any time $t$, where $\Omega$ is
the sample space of the probability space $(\Omega, {\mathcal F},
{\mathbb P})$ on which the stochastic dynamics is defined; as a
matter of fact, at the end of a measurement process we know that the
pointer is located somewhere along the scale, but we do not know
exactly where. Accordingly, the physically relevant information is
embodied in the {\it probability distribution} of~$q_{t}$:
\begin{equation} \label{eq:prob-distr}
{\mathbb P}_{q_{t}}[B] \quad \equiv \quad {\mathbb P}[q_{t}^{-1}(B)]
\qquad\quad B \, \in \, {\mathcal B({\mathbb R})},
\end{equation}
which gives the probability that~$q_{t}$ lies within the
(measurable) subset $B$ of the Borel $\sigma$-algebra ${\mathcal
B({\mathbb R})}$ of ${\mathbb R}$.

\subsubsection{The calibration of an experiment}
\label{sec:hsop-drm:experiment-calibration}

As we mentioned at the beginning of this section, we  usually do not
speak of the outcome of an experiment in terms of ``the pointer
being in a particular position in space'', but rather in terms of
``the pointer signaling a particular outcome'', typically a real
number; this sort of association between the spatial position of the
pointer and the (numerical) outcome of the experiment is of course
entirely a matter of convention. E.g., in a Stern-Gerlach type of
experiment we usually associate the upper part of the screen with
the outcome~$+1$ and the lower part with the outcome~$-1$, but we do
not directly observe the number $+1$ or $-1$: what we observe are
spots either in the upper or in the lower part of the plate or, if
we want to keep referring to a movable pointer, what we observe is
the pointer sitting in two macroscopically different positions along
the scale, corresponding to the two possible outcomes. The set of
outcomes $\set{-1, +1}$ is conventional, and we are free to use
whatever set suits best the interpretation of the experiment. In
accordance with Ref.~\cite{bm}, in what follows we will refer to the
association between directly observed positions of the pointer and
conventionally decided outcomes as to the \emph{calibration} of the
experiment. In all generality, we define
the calibration of an experiment as follows. \\

\noindent \textsc{Definition 2: Calibration function}. Let
${\mathcal O}$ be the set of possible outcomes of an experiment. The
function
\begin{equation} \label{eq:293}
f : \; \left\{
\begin{array}{lcl}
{\mathbb R} & \rightarrow & {\mathcal O} \\
q & \rightarrow & f(q)
\end{array}
\right.
\end{equation}
from the space~${\mathbb R}$ of the positions of (the center of mass
of) the pointer to the space~$\mathcal{O}$ of the possible outcomes
of the experiment, is called the \emph{calibration function} of the
experiment. For mathematical convenience, we assume~$f$ to be
${\mathcal B}({\mathbb R})/{\mathcal G}$-measurable, where
${\mathcal G}$ is a chosen
$\sigma$-algebra on ${\mathcal O}$. \\

\noindent By means of the calibration function, we can replace the
probability distribution ${\mathbb P}_{q_{t}}$ of
Eq.~\eqref{eq:prob-distr} which gives the probability that the
pointer lies within $B \in {\mathbb R}$, with the probability
distribution
\begin{equation}
{\mathbb P}_{f(q_{t})}[V] \quad \equiv \quad {\mathbb P}[
q^{-1}_{t}(f^{-1}(V))] \qquad\quad V \, \in \, {\mathcal G},
\end{equation}
which represents the probability that the outcome of the experiment
belongs to the measurable subset $V$ of ${\mathcal O}$, the set of
all possible outcomes.

\subsection{The link between experimental outcomes and microscopic states}
\label{sec:hsop-drm:outcomes_microstates-link}

In this section we establish the main result of the paper, namely
the link between the probabilities associated to the possible
outcomes of an experiment and the pre-measurement states of the
microscopic system whose properties are measured, showing how such
probabilities can be represented by the mean value of the effects of
a POVM over the state of the microscopic system right before the
measurement.

\subsubsection{The characteristic traits of measurement-like situations}
\label{sec:hsop-drm:experiment-traits}

We start our analysis by summarizing the most relevant properties of
measurement processes, as they are described within DRMs; for a
detailed analysis of these properties, we refer the reader to
Ref.~\cite{mis}.

Let $x_{n}$ be the $n$-th position of the pointer along the graduate
scale corresponding to the outcome $v_{n} \in {\mathcal O}$, where
${\mathcal O}$ is the set of all possible outcomes, to which a
$\sigma$-algebra ${\mathcal G}$ is associated. Let
$t_{\makebox{\tiny F}}$ be the time at which the experiment ends.
According to the dynamical laws typical of all DRMs, we can state
that:
\begin{description}
\item[\it Property 1.]
Throughout the whole measurement process, and in particular at its
end, the center of mass of the pointer is always very well localized
in space; making reference to the model analyzed in Ref.~\cite{mis},
the spread in position of the center-of-mass wave function of a
pointer having the mass of 1 g is about $\sigma_{q} \simeq 10^{-14}$
m.

\item[\it Property 2.]
Let us consider an interval $\delta_{n}$, centered around the
position $x_{n}$, whose extension $d$ we take of the order of
$10^{-5}$ cm; then, the probability $\eta$ that, at the end of the
measurement process, $q_{t_{\makebox{\tiny F}}}$ does not lie inside
{\it any} of the sub-intervals $\delta_{n}$ is very small, e.g.
$\eta \simeq 10^{-17}$. This means that, with probability extremely
close to 1, the pointer ends up in one of the positions along the
graduate scale corresponding to one of the possible outcomes.
\end{description}

Since the wave function of the center of mass of the pointer has a
spatial extension (though very small), the event
$q_{t_{\makebox{\tiny F}}} \in \delta_{n}$ does not imply that most
of the center-of-mass wave function lies almost entirely within
$\delta_{n}$: this happens e.g. when $q_{t_{\makebox{\tiny F}}}$
lies at the border of $\delta_{n}$. In order to take this
possibility properly into account, let us consider a new set of
intervals $\Delta_{n}$ containing the intervals $\delta_{n}$, each
of which is centered around $x_{n}$, whose extension is equal to $d
+ \ell$ where $\ell$ is large compared to the typical spread
$\sigma_{q}$ of the center of mass of the pointer, e.g. $\ell \simeq
10^{8} \times \sigma_{q} \simeq 10^{-6}$ m (see Fig.~\ref{fig1}). We
can then say that when $q_{t_{\makebox{\tiny F}}}$ lies within
$\delta_{n}$, practically all the center-of-mass wave function lies
within $\Delta_{n}$, while the probability for $q_{t_{\makebox{\tiny
F}}}$ lying in $\Delta_{n}-\delta_{n}$, i.e. outside $\delta_{n}$,
is, according to property 2, vanishingly small.
\begin{figure}
\begin{center}
{\includegraphics[scale=0.7]{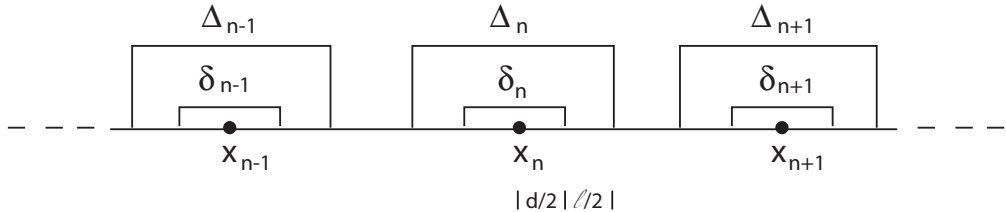}} \caption{The picture shows
the points $x_{n}$ associated to the different outcomes of an
experiment, together with the intervals $\delta_{n}$ and
$\Delta_{n}$ used for the proof of Theorem 1.} \label{fig1}
\end{center}
\end{figure}

Up to now we have spoken only of the position of the pointer, but as
we have said it is custom to refer the probabilities to the
(numerical) outcomes of the measurement. To this end we introduce
the following calibration function:
\begin{equation}
f(q_{t_{\makebox{\tiny F}}}) \; = \; v_{n} \, \in \, {\mathcal O}
\qquad\quad \text{iff:}\;\; q_{t_{\makebox{\tiny F}}} \, \in \,
\Delta_{n},
\end{equation}
which simply means that the outcome of the experiment is $v_{n}$
whenever the pointer, at the end of the measurement process, lies
around the position $x_{n}$ of the graduate scale. In the above
definition, we have chosen $\Delta_{n}$ as the relevant intervals in
order to make the proof of the following theorem simpler: however,
on a macroscopic scale the two sets of intervals ($\{\delta_{n}\}$
and $\{ \Delta_{n} \}$) are practically identical. Note that, since
we want different outcomes to correspond to different macroscopic
configurations of the pointer, the intervals $\Delta_{n}$ should not
overlap, which means that the distance between two consecutive
points $x_{n}$ and $x_{n+1}$ along the graduate scale should be
bigger that $d + \ell \simeq 10^{-6}$ m; this is of course a
perfectly reasonable assumption.

We are now in the position to prove the following theorem, which is
the cornerstone for the subsequent derivation of the Hilbert space
formalism from the framework of DRMs.\\

\noindent \textsc{Theorem 1.} Let us consider an experiment globally
described by the stave vector $|\psi_{t}\rangle$, let $q_{t}$ be the
position of the (center of mass of the) pointer at time $t$ right
before the measurement begins, and let $t_{\makebox{\tiny F}}$ be
the time at which the experiment ends. Then:
\begin{equation} \label{eq:306}
\forall \,\, V \, \in \, {\mathcal G}: \qquad \left|{\mathbb
P}_{f(q_{t_{\makebox{\tiny F}}})}[V] - {\mathbb
Q}_{t_{\makebox{\tiny F}}}[V] \right| \; \leq \; 2 \left[
\left(\frac{\sigma_{q}}{\ell} \right)^{2} + \eta \right],
\end{equation}
where the probability measure ${\mathbb Q}_{t}$ is defined as
follows:
\begin{equation} \label{eq:npm}
{\mathbb Q}_{t}[V] \; \equiv \; {\mathbb E}_{\mathbb P} [\langle
\psi_{t} | P_{\Delta_{V}} | \psi_{t}\rangle ],
\end{equation}
and $P_{\Delta_{V}}$ is the projection operator on the (measurable)
subset $\Delta_{V} \equiv f^{-1}(V)$, which is the subset of
${\mathbb R}$ of all positions of the pointer corresponding to a
precise outcome $v$ among the possible outcomes belonging to $V$.
\\

\noindent The above theorem is very powerful: according to the very
general structures of DRMs, the probability distribution ${\mathbb
P}_{f(q_{t_{\makebox{\tiny F}}})}$ of the outcomes of an experiment
can, for all practical purposes, be replaced by the probability
measure ${\mathbb Q}_{t_{\makebox{\tiny F}}}$ which has a the very
simple mathematical expression given by Eq.~\eqref{eq:npm}. This is
due to Eq.~\eqref{eq:306} and to the fact that, according to the
typical numerical values for $\lambda, \eta$, $\sigma_q$ and $\ell$
given before, one has:
\begin{equation}
2 \left[ \left(\frac{\sigma_{q}}{\ell} \right)^{2} + \eta \right] \;
\simeq \; 2 \times 10^{-16},
\end{equation}
a vanishingly small value. We point out that Eq.~\eqref{eq:npm}
should not be confused with the quantum average value of the
projection operator $P_{B}$ relative to the position of the pointer,
as given by standard quantum mechanics. As a matter of fact, since
an experiment always involves a macroscopic system, the one
described by $|\psi_{t}\rangle$, a state vector which includes both
the measured system and the measuring apparatus entails a dynamical
evolution which is entirely different from that predicted by the
Schr\"odinger equation, in particular it always keeps the measuring
device in a state well localized in space. We now prove the theorem.
\begin{eqnarray} \label{eq:298}
\left|{\mathbb P}_{f(q_{t_{\makebox{\tiny F}}})}[V] - {\mathbb
Q}_{t_{\makebox{\tiny F}}}[V] \right| & \equiv & \left|{\mathbb
P}[q_{t_{\makebox{\tiny F}}}^{-1}(\Delta_{V})] \; - \; {\mathbb
E}_{\mathbb P} [\langle \psi_{t_{\makebox{\tiny F}}} |
P_{\Delta_{V}} |
\psi_{t_{\makebox{\tiny F}}} \rangle] \right| \nonumber \\
& = & \left| \int_{q_{t_{\makebox{\tiny F}}}^{-1}(\Delta_{V})} d
{\mathbb P} \; - \; \int_{\Omega} \langle \psi_{t_{\makebox{\tiny
F}}} | P_{\Delta_{V}} | \psi_{t_{\makebox{\tiny F}}} \rangle \, d
{\mathbb
P} \right| \nonumber \\
& \leq & \int_{q_{t_{\makebox{\tiny F}}}^{-1}(\Delta_{V})} ( 1 -
\langle \psi_{t_{\makebox{\tiny F}}} | P_{\Delta_{V}} |
\psi_{t_{\makebox{\tiny F}}} \rangle ) d {\mathbb P} \; + \;
\int_{\Omega - q_{t_{\makebox{\tiny F}}}^{-1}(\Delta_{V})} \langle
\psi_{t_{\makebox{\tiny F}}} | P_{\Delta_{V}} |
\psi_{t_{\makebox{\tiny F}}} \rangle \, d {\mathbb P} \nonumber
\\
& = & \int_{q_{t_{\makebox{\tiny F}}}^{-1}(\Delta_{V})} \langle
\psi_{t_{\makebox{\tiny F}}} | P_{{\mathbb R} - \Delta_{V}} |
\psi_{t_{\makebox{\tiny F}}} \rangle  d {\mathbb P} \; + \;
\int_{q_{t_{\makebox{\tiny F}}}^{-1}({\mathbb R} - \Delta_{V})}
\langle \psi_{t_{\makebox{\tiny F}}} | P_{\Delta_{V}} |
\psi_{t_{\makebox{\tiny F}}} \rangle \, d {\mathbb P},
\end{eqnarray}
where $\Omega$ is the sample space on which the stochastic dynamics
is defined. Let us define $\delta_{V} \equiv \cup_{n} \delta_{n}$
with $\delta_{n} \, \in \, \Delta_{V}$; the first of the two terms
in the last line can be re-written as follows:
\begin{equation} \label{eq:nuo}
\int_{q_{t_{\makebox{\tiny F}}}^{-1}(\Delta_{V})} \langle
\psi_{t_{\makebox{\tiny F}}} | P_{{\mathbb R} - \Delta_{V}} |
\psi_{t_{\makebox{\tiny F}}} \rangle \, d {\mathbb P} \; = \;
\int_{q_{t_{\makebox{\tiny F}}}^{-1}(\delta_{V})} \langle
\psi_{t_{\makebox{\tiny F}}} | P_{{\mathbb R} - \Delta_{V}} |
\psi_{t_{\makebox{\tiny F}}} \rangle \, d {\mathbb P} \; + \;
\int_{q_{t_{\makebox{\tiny F}}}^{-1}(\Delta_{V} - \delta_{V})}
\langle \psi_{t_{\makebox{\tiny F}}} | P_{{\mathbb R} - \Delta_{V}}
| \psi_{t_{\makebox{\tiny F}}} \rangle \, d {\mathbb P};
\end{equation}
according to the second property of DRMs listed above, the
probability measure of the set $\Delta_{V} - \delta_{V}$ is smaller
than $\eta$; thus, taking also into account that $\langle
\psi_{t_{\makebox{\tiny F}}} | P_{{\mathbb R} - \Delta_{V}} |
\psi_{t_{\makebox{\tiny F}}} \rangle \leq 1$, we can write:
\begin{equation} \label{eq:bou1}
\int_{q_{t_{\makebox{\tiny F}}}^{-1}(\Delta_{V} - \delta_{V})}
\langle \psi_{t_{\makebox{\tiny F}}} | P_{{\mathbb R} - \Delta_{V}}
| \psi_{t_{\makebox{\tiny F}}} \rangle \, d {\mathbb P} \; \leq \;
\eta.
\end{equation}
Regarding the first term on the right-hand-side of
Eq.~\eqref{eq:nuo}, we now show that the integrand $\langle
\psi_{t_{\makebox{\tiny F}}} | P_{{\mathbb R} - \Delta_{V}} |
\psi_{t_{\makebox{\tiny F}}} \rangle$ is extremely small for any
$\omega \in \Omega$ such that~$q_{t_{\makebox{\tiny F}}}$ lies
within $\delta_{V}$. Defining $\psi(x) = \langle x |
\psi_{t_{\makebox{\tiny F}}}\rangle$, we have of course
\begin{equation} \label{eq:301}
\langle \psi_{t_{\makebox{\tiny F}}} | P_{{\mathbb R} - \Delta_{V}}
| \psi_{t_{\makebox{\tiny F}}} \rangle = \int d \{ q \}
\int_{{\mathbb R} - \Delta_{V}} dx\; |\psi_{t_{\makebox{\tiny
F}}}(\{q\},x)|^{2},
\end{equation}
where $\{ q \}$ denote all degrees of freedom involved in the
measurement, except the one for the center of mass of the pointer.
Whenever~$q_{t_{\makebox{\tiny F}}}$ belongs to $\delta_{V}$, we
have that the parameter $\ell$ defining essentially the extension of
the intervals $\Delta_{n}$, satisfies $\ell^{2} \leq (x -
q_{t_{\makebox{\tiny F}}})^{2}$ for all~$x \in {\mathbb R} -
\Delta_{V}$; we get then the following inequality:
\begin{eqnarray} \label{eq:302}
\ell^{2} \int d \{ q \} \int_{{\mathbb R} - \Delta_{V}} dx\;
|\psi_{t_{\makebox{\tiny F}}}(\{q\},x)|^{2} & \leq &  \int d \{ q
\}\int_{{\mathbb R} - \Delta_{V}} dx\; (x - q_{t_{\makebox{\tiny
F}}})^2 |\psi_{t_{\makebox{\tiny
F}}}(\{q\},x)|^{2} \nonumber \\
& \leq & \int d \{ q \} \int_{{\mathbb R}} dx\; (x -
q_{t_{\makebox{\tiny F}}})^2 |\psi_{t_{\makebox{\tiny
F}}}(\{q\},x)|^{2} \; \equiv \; \sigma_{q}^{2}.
\end{eqnarray}
Eqs.~\eqref{eq:nuo}, \eqref{eq:bou1} and~\eqref{eq:302} lead to the
following result:
\begin{equation}
\int_{q_{t_{\makebox{\tiny F}}}^{-1}(\Delta_{V})} \langle
\psi_{t_{\makebox{\tiny F}}} | P_{{\mathbb R} - \Delta_{V}} |
\psi_{t_{\makebox{\tiny F}}} \rangle  d {\mathbb P} \; \leq \;
\left( \frac{\sigma_{q}}{\ell} \right)^2 + \eta
\end{equation}
A completely symmetric argument holds for the second and last terms
at the right-hand-side of equation~\eqref{eq:298} as well, hence the
theorem is proven.

To summarize, we have shown that in measurement-like situations it
is fully legitimate, as a consequence of the reducing dynamics of
DRM, to use the probability measure~${\mathbb Q}_{t}$ in place of
${\mathbb P}_{f(q_{t})}$ to compute the probabilities of the
possible outcomes of an experiment.

\subsubsection{The emergence of the Hilbert-space operator formalism}
\label{sec:hsop-drm:hsop-emergence}

Let us now focus our attention on the probability measure ${\mathbb
Q}_{t}$ which, according to the definition~\eqref{eq:npm}, can be
re-written also as:
\begin{equation} \label{eq:4556}
{\mathbb Q}_{t}[V] \; \equiv \; \makebox{Tr}\,[\rho_{t} \, ({\mathbb
I}_{\{q\}} \otimes P_{\Delta_{V}}) ] \qquad \quad V \, \in \,
{\mathcal G},
\end{equation}
where ${\mathbb I}_{\{q\}}$ is the identity operator acting on the
space of all degrees of freedom of the experiment, except the one
referring to the center of mass of the pointer. The above expression
is a consequence of the well-known formula typical of
DRMs~\cite{rev1}:
\begin{equation}
{\mathbb E}_{\mathbb P}[ \langle \psi_{t} | O | \psi_{t} \rangle] \;
= \; \makebox{Tr}\, [ O \rho_{t}] \qquad\quad \text{with: $\rho_{t}
\equiv {\mathbb E}_{\mathbb P}[ |\psi_{t}\rangle\langle \psi_{t}
|]$},
\end{equation}
where $O$ is any suitable operator. We remind the reader that the
state vector $|\psi_{t}\rangle$ (thus also $\rho_{t}$) refers both
to the (whole) state of the measuring device as well as to the state
of the microscopic system. Regarding the initial state
$|\psi_{t_{0}}\rangle$ (or, equivalently, $\rho_{t_{0}}$), right
before the experiment begins, we make the following assumption:
\begin{description}
\item[\it Assumption 1.]
At the beginning of the experiment ($t = t_{0}$), the state of the
micro-system and that of the apparatus are factorized:
\begin{equation}
\qquad\rho_{t_{0}} \; = \; \rho^{S}_{t_{0}} \otimes
\rho^{A}_{t_{0}},
\end{equation}
where $\rho^{S}_{t_{0}} =  |\psi^{S}_{t_{0}}\rangle\langle
\psi^{S}_{t_{0}}|$ represents the initial state of the microscopic
system, which we assume to be pure, while $\rho^{A}_{t_{0}}$
represents the initial state of the apparatus.
\end{description}
This initial factorization of the two states is a very natural
assumption to make, since otherwise the microscopic system would not
be in any defined state, whose property the experiment should
detect. Moreover, we recall the following important property
characterizing dynamical reduction models:
\begin{description}
\item[\it Property 3.]
The dynamical evolution $\Sigma_{(t_{0},t_{\makebox{\tiny F}})}$
mapping the density matrix $\rho_{t_{0}}$ describing the global
state prior to the experiment, into the final state
$\rho_{t_{\makebox{\tiny F}}}$ after the experiment, is of the
quantum dynamical semigroup type, thus, in particular, linear and
trace-preserving.
\end{description}
According to the above two assumptions, we can write ${\mathbb
Q}_{t}$, computed at time $t = t_{\makebox{\tiny F}}$, as follows:
\begin{equation} \label{eq:fin}
{\mathbb Q}_{t_{\makebox{\tiny F}}}[V] \; \equiv \;
\makebox{Tr}\,[\Sigma_{(t_{0},t_{\makebox{\tiny F}})}
(\rho^{S}_{t_{0}} \otimes \rho^{A}_{t_{0}}) {\mathbb I}_{\{q\}}
\otimes P_{\Delta_{V}} ] \qquad \quad V \, \in \, {\mathcal G},
\end{equation}
for any fixed $V \in {\mathcal G}$. Assuming now that
$|\psi^{S}_{t_{0}}\rangle$ can run over the entire Hilbert space
${\mathcal H}^{S}$, the above expression defines the diagonal part
of the following bounded sesquilinear form:
\begin{equation}
\begin{array}{ccl}
{\mathcal H}^{S} \otimes {\mathcal H}^{S} & \longrightarrow &
{\mathbb C} \\
(|\psi^{S}_{t_{0}}\rangle, |\phi^{S}_{t_{0}}\rangle) &
\longrightarrow & \makebox{Tr}\,[\Sigma_{(t_{0},t_{\makebox{\tiny
F}})} (|\psi^{S}_{t_{0}}\rangle\langle \phi^{S}_{t_{0}}| \otimes
\rho^{A}_{t_{0}}) {\mathbb I}_{S} \otimes P_{\Delta_{V}} ];
\end{array}
\end{equation}
which, according to the Riesz representation theorem, can be written
as follows:
\begin{equation} \label{eq:drf}
{\mathbb Q}_{t_{\makebox{\tiny F}}}[V] \quad = \quad \langle
\psi^{S}_{t_{0}} | O_{V} | \psi^{S}_{t_{0}} \rangle \quad = \quad
\text{Tr}\, [\rho_{t_{0}} O_{V} ],
\end{equation}
where $O_{V}$ is a bounded linear operator in ${\mathcal H^{S}}$. In
our case, $O_{V}$ turns out to be also self-adjoint and defines a
POVM from the measurable space of the possible outcomes $({\mathcal
O}, {\mathcal G})$ to the Hilbert space ${\mathcal H}^{S}$ of the
micro-system. This is the desired result, which
we formalize in the following theorem: \\

\noindent \textsc{Theorem 2.} According to the properties of DRMs
stated before (properties 1--3), according to assumption 1 and,
within the limits set by Theorem 1, one can write:
\begin{equation}
{\mathbb P}_{f(q_{t_{\makebox{\tiny F}}})}[V] \quad \simeq \quad
\langle \psi^{S}_{t_{0}} | O_{V} | \psi^{S}_{t_{0}} \rangle
\qquad\quad \forall \,\, V \, \in \, {\mathcal G}.
\end{equation}
In other words, the probability that the outcome of a given
experiment belongs to the measurable subset $V$ of the set
${\mathcal O}$ of possible outcomes can be (with very high accuracy)
expressed as the average value of the effect $O_{V}, V \in {\mathcal
G}$, of a POVM, over the initial
state $| \psi^{S}_{t_{0}} \rangle$ of the microscopic system. \\

\noindent We have thus recovered the operator formalism of standard
quantum mechanics.

It is interesting to compare Eq.~\eqref{eq:drf} with
Eq.~\eqref{eq:fin}:
\begin{equation}  \label{eq:319}
{\mathbb Q}_{t_{\makebox{\tiny F}}}[V] \quad = \quad
\makebox{Tr}\,[\Sigma_{(t_{0},t_{\makebox{\tiny F}})}
(\rho^{S}_{t_{0}} \otimes \rho^{A}_{t_{0}}) {\mathbb I}_{S} \otimes
P_{\Delta_{V}} ] \quad = \quad \text{Tr}\, [\rho_{t_{0}} O_{V} ];
\end{equation}
the middle term of the above equation provides the true physical
description of the experiment: it gives the probability for the
pointer to lie within a well-defined region along the graduate scale
at time $t_{\makebox{\tiny F}}$; on the other hand, the last term
provides the compact (and very handy) quantum way of expressing such
probabilities in terms of the initial state of the micro-system. A
couple of further comments are at order.

\noindent 1.  Clearly, since a wave function has always a spatial
extension, the width of the intervals $\Delta_{n}$ can not be set
equal to zero. This means that, according to DRMs, only experiments
having at most a countable number of outcomes can be performed. This
of course includes all physically realizable experiments, while
those having a continuous number of outcomes represent a
mathematical idealization.

\noindent 2. Nowhere in our analysis we have explicitly used the
fact that the probabilities of the possible outcomes of the
experiment, as predicted by DRMs, practically coincides with quantum
probabilities. However, such a feature of DRMs is implicitly
contained in property 3, i.e. in the fact that the evolution law for
the statistical operator is linear; the reason is the following.
When a jump process of the form~\eqref{eq:sdfsd} occurs on the
$n$-th particle of a many-particle system, a density matrix $\rho
\equiv \sum_{i} c_{i} |\psi_{i}\rangle\langle\psi_{i}|$ changes, in
accordance to axiom B, as follows:
\begin{equation}
\rho \equiv \sum_{i} c_{i} |\psi_{i}\rangle\langle\psi_{i}| \quad
\longrightarrow \quad \sum_{i} c_{i} \int d^3 {\bf x}\; p({\bf x})\;
\frac{L_{n}({\bf x})|\psi_{i}\rangle\langle\psi_{i}|L_{n}({\bf
x})}{\| L_{n}({\bf x}) \psi_{i}\rangle \|^2},
\end{equation}
where $p({\bf x})$ is the probability density for a jump to occur in
${\bf x}$. As we see, the above evolution is {\it not} linear in
$\rho$, unless we require $p({\bf x})$ to be of the
form~\eqref{eq:ghhfd}, which agrees with the Born probability rule.

\section{Reproducibility and PVM} \label{sec:hsof:reproducibility}

The example we have discussed in Sec.~\ref{sec:hsof:example} belongs
to a particular subclass of experiments, because it can be
associated to a PVM, while the general theorem of the previous
section shows that experiments are in general associated to POVM,
which are more general than PVM. In this section we show that the
possibility of associating a PVM to an experiment is strictly
connected to the \emph{reproducibility} of the experiment itself:
since on standard books on quantum mechanics it is (implicitly)
assumed that experiments are reproducible, the analysis shows why
usually experiments are associated to projection operators.

We say that an experiment is reproducible if it can be performed
many times on the same physical system and, each time we perform two
runs in a row, one right after the other, the second one always
yields the same outcome as the first one. In order to give a more
rigorous definition, we first have to recall another important
feature of DRMs:
\begin{description}
\item[\it Property 4.]
At the end of a measurement process, the final state
$|\psi_{t_{\makebox{\tiny F}}}\rangle$ of the whole system is, to a
great accuracy, a factorized state of the system and the
apparatus\footnote{See~\cite{mis} for a quantitative analysis of
this feature of DRMs}: $|\psi_{t_{\makebox{\tiny F}}}\rangle \simeq
|\psi_{t_{\makebox{\tiny F}}}^{S}\rangle \otimes
|\psi_{t_{\makebox{\tiny F}}}^{A}\rangle$. Accordingly, the vector
$|\psi^{S}_{t_{0}}\rangle$ representing the state of the microscopic
system right before the experiment began, changes to a new
(normalized) state $|\psi^{S}_{t_{\makebox{\tiny F}}}\rangle$, which
depends in general both on $|\psi^{S}_{t_{0}}\rangle$ and on the
stochastic dynamics of the interaction between the system $S$ and
the apparatus $A$.
\end{description}
In the following, when convenient we will write
$|\psi^{S}_{t_{\makebox{\tiny F}}}/\psi^{S}_{t_{0}}\rangle$ in place
of $|\psi^{S}_{t_{\makebox{\tiny F}}}\rangle$, to stress the
dependence of the final state on the initial one; moreover, we will
often write $|\psi^{S,v_{n}}_{t_{\makebox{\tiny F}}}\rangle$ in
place of $|\psi^{S}_{t_{\makebox{\tiny F}}}\rangle$ to signify that
$|\psi^{S}_{t_{\makebox{\tiny F}}}\rangle$ is the state to which the
systems $S$ is reduced at the end of the first experiment in which
we suppose that the outcome $v_{n}$ has been obtained. Let
$|\psi^{2:v_{n}}_{t}\rangle$ describe the time evolution during the
second experiment, when the initial state at time $t_{0}'$ at which
the second experiment takes place is assumed to be:
$|\psi^{2:v_{n}}_{t_{0}'}\rangle =
|\psi^{S,v_{n}}_{t_{\makebox{\tiny F}}}\rangle \otimes
|\psi^{A}_{t_{0}'}\rangle$.
A reproducible experiment is defined as follows.\\

\noindent \textsc{Definition 3: Reproducible experiment.} An
experiment on a physical system $S$ is said to be
\emph{reproducible} if and only if:
\begin{enumerate}
\item\label{item:34}
The experiment can be performed on $S$ at least twice; moreover, the
Hilbert space ${\mathcal H}$ of vectors describing the possible
states of $S$ before the measurement is the same as the Hilbert
space of vectors describing the possible states of the system after
the measurement; in other words, the totality of the possible final
states $|\psi^{S}_{t_{\makebox{\tiny F}}}\rangle$ must span
${\mathcal H}$:
\begin{equation} \label{eq:122}
\linearspan \set{|\psi^{S}_{t_{\makebox{\tiny
F}}}/\psi^{S}_{t_{0}}\rangle, \;\; |\psi^{S}_{t_{0}}\rangle \, \in
\, {\mathcal H}} = \mathcal{H}.
\end{equation}
\item\label{item:35}
Let us suppose that two experiments are done, one immediately after
the other; let ${\mathbb P}^{v_{n}}_{f(q_{t_{\makebox{\tiny F}}'})}$
be the probability distribution of the outcomes of the second
experiment, assuming that $|\psi^{S:v_{n}}_{t_{\makebox{\tiny
F}}}\rangle$ has been taken as the initial state of the microscopic
system for the second experiment, i.e. assuming that the outcome of
the first experiment is $v_{n}$. We then require that:
\begin{equation} \label{eq:121}
{\mathbb P}^{v_{n}}_{f(q_{t_{\makebox{\tiny F}}'})}[V] = 1, \qquad
\forall \,\, V\, : v_{n} \in V
\end{equation}
i.e. that the outcome of the second run of the experiment belongs
to~$V$ with certainty.
\end{enumerate}

\noindent Let us briefly comment on the above definition. The first
request in the definition above excludes experiments which alter the
nature of the physical system, e.g. because they destroy it, or one
of its parts, or because they transform it in a new physical system
with a space of states~$\mathcal{H}'$ different from the original
space~$\mathcal{H}$. The second request just embodies the idea of
reproducibility, by assuring that the same outcome is obtained with
certainty when the experiment is performed twice.
Now we can state the following theorem: \\

\noindent \textsc{Theorem 3: Reproducible experiment.} Let us
consider an experiment which, according to Theorem 2, is associated
to a POVM $\{ O_{V} \}_{V}$. If the experiment is reproducible, and
within the limits of replacing the probability ${\mathbb
P}^{v_{n}}_{f(q_{t_{\makebox{\tiny F}}'})}$ with the probability
${\mathbb Q}^{v_{n}}_{t_{\makebox{\tiny F}}'}[V] = {\mathbb
E}_{\mathbb P} [ \langle \psi^{2:v_{n}}_{t_{\makebox{\tiny F}}'} |
P_{\Delta_{V}} |\psi^{2:v_{n}}_{t_{\makebox{\tiny F}}'}\rangle
]$, then the POVM is a PVM. \\

\noindent The proof of the theorem is given in Ref.~\cite{bm}; here
we propose a simplified version of it. We first of all notice that,
given a self-adjoint positive semidefinite operator $O$ with bounds
$0 \leq O \leq I$, then
\begin{eqnarray}
\langle \psi | O |\psi\rangle = 1 & \qquad \Rightarrow \qquad & O
|\psi\rangle =
|\psi\rangle \label{eq:prop1} \\
\langle \psi | O |\psi\rangle = 0 & \Rightarrow & O |\psi\rangle =
|\omega\rangle,  \label{eq:prop2}
\end{eqnarray}
where $|\omega\rangle$ is the null vector. To show this, let us
write ${\mathcal H} = \mathcal{H}_{1} \oplus
\mathcal{H}_{1}^{\perp}$, where $\mathcal{H}_{1}$ is the subspace of
all eigenstates of $O$ corresponding to the eingenvalue 1 and
$\mathcal{H}_{1}^{\perp}$ its orthogonal complement. Let
$|\psi\rangle$ be a normalized vector, which we decompose in:
$\ket{\psi} = \ket{\psi_{\parallel}} + \ket{\psi_{\perp}}$, where
$\ket{\psi_{\parallel}} \in \mathcal{H}_{1}$ and $\ket{\psi_{\perp}}
\in \mathcal{H}_{1}^{\perp}$. It follows that:
\begin{equation} \label{eq:126}
\langle \psi | O |\psi\rangle = 1 \qquad \Rightarrow \qquad
\Braket{\psi_{\perp} | O | \psi_{\perp}} = \Braket{\psi_{\perp} |
\psi_{\perp}}.
\end{equation}

On the other hand, the property $0 \leq O \leq I$ implies
that~$O$~acts on the subspace~$\mathcal{H}_{1}^{\perp}$ like a
contraction, i.e. that $\| O \ket{\psi_{\perp}} \| < \|
\ket{\psi_{\perp}} \|$, unless $\ket{\psi_{\perp}} =
|\omega\rangle$. This contraction property, together with the
Cauchy-Schwarz inequality, gives
\begin{equation} \label{eq:131}
\braket{\psi_{\perp} | O | \psi_{\perp}} \; < \;
\braket{\psi_{\perp} | \psi_{\perp}} \qquad \text{unless} \;\;
\ket{\psi_{\perp}} = |\omega\rangle.
\end{equation}
Now, Eqs.~\eqref{eq:126} and~\eqref{eq:131} are incompatible unless
$\ket{\psi_{\perp}} = |\omega\rangle$, which proves
Eq.~\eqref{eq:prop1}. In a similar way one can prove also
Eq.~\eqref{eq:prop2}.

Coming back to the probability measure ${\mathbb
Q}^{v_{n}}_{t_{\makebox{\tiny F}}'}$, according to the analysis of
the previous section and to the hypotheses of our theorem, we can
write, for any $V \subseteq {\mathcal G}$:
\begin{equation} \label{eq:fdbd}
{\mathbb Q}^{v_{n}}_{t_{\makebox{\tiny F}}'}[V] \; = \; \langle
\psi^{S:v_{n}}_{t_{\makebox{\tiny F}}} | O_{V} |
\psi^{S:v_{n}}_{t_{\makebox{\tiny F}}} \rangle \; = \; \left\{
\begin{array}{ll}
1 \quad & \text{if} \; v_{n} \in V, \\
0 & \text{else.}
\end{array}
\right.
\end{equation}
Let ${\mathcal H}_{1}^{V}$ be the subspace of all eigenstates of
$O_{V}$ corresponding to the eingenvalue 1, and ${\mathcal
H}_{0}^{V}$ be the subspace of all eigenstates corresponding to the
eingenvalue 0; according to~\eqref{eq:fdbd}, and to the fact that $
\text{span} \{ | \psi^{S:v_{n}}_{t_{\makebox{\tiny F}}} \rangle \}
\equiv {\mathcal H}$, we can write: ${\mathcal H} = {\mathcal
H}_{1}^{V} \oplus {\mathcal H}_{0}^{V}$. This result proves
that~$O_{V}$ is an orthogonal projector on $\mathcal{H}$, with
support in~${\mathcal H}_{1}^{V}$.  Hence the POVM $\set{O_{V}}_{V}$
is a PVM.

The \emph{ideal measurements} usually considered in many QM
textbook, i.e., those obeying the rule of the Wave-Packet
Reduction~(WPR) postulate, are reproducible, and therefore the POVM
associated to them reduces to a PVM. When this happens, the
experiment can be entirely characterized, with respect to its
statistical properties, by a single self-adjoint operator, as is
usually done in quantum mechanics.

\section{The ``tail problem''; the Stern-Gerlach experiment
revisited} \label{sec:hsop-drm:stern_gerlach_experiment}

One of the reason why, with reference to theorem 1, the two
probabilities ${\mathbb P}_{q_{t_{\makebox{\tiny F}}}}$ and
${\mathbb Q}_{t_{\makebox{\tiny F}}}$ are not strictly equal is
that, even when $q_{t} \in \delta_{n}$, the wave function for the
center of mass of the pointer has not a compact support contained in
the interval $\Delta_{n}$, but it has tails spreading out to
infinity; such tails, being extremely small (in the sense that the
integral of the square modulus of the wave function over the region
laying outside $\Delta_{n}$, assuming that its center is contained
within $\delta_{n}$, is very small), give not rise to any problem in
connection with the interpretation of the theory and its physical
predictions.

In this section we want to point out that precisely the same problem
with tails occur also in standard quantum mechanics. Let us take as
an example the Stern-Gerlach experiment: this is often used in
textbooks as \emph{the} paradigmatic experiment which illustrates
the correspondence between observables and self-adjoint operators.
What textbooks usually provide is just a simplified description of
the truly observed experimental results; a more realistic analysis
which takes into account also the spatial (beside spin) degrees of
freedom of the atoms sent through the apparatus would show that
tails emerge also here, which (just in principle) give rise to some
potential problems in interpreting the outcome of the experiment,
and in associating a self-adjoint operator to it. In this section we
perform such a kind of analysis, we discuss the role of the tails of
the wave function of the atoms and we make precise the sense in
which it is legitimate to associate the usual spin operator~$S_{z}$
to the experiment.

As it is well known, Stern and Gerlach used an oven to produce and
send a beam of silver atoms through an inhomogeneous magnetic field,
letting it eventually impinge on a glass plate. In order to analyze
the effect of the magnetic field on the beam, two separate
experiments were originally \cite{sg} performed: one with the magnet
generating the field turned on, with a run time of~8 hours, another
with the magnet turned off, with a run time of~4.5 hours. In the
magnet-off case, a single bar of silver approximately 1.1 mm long
and 0.06--0.1 mm wide was deposited on the glass plate. In the
magnet-on case, a pair-of-lips shape appeared on the glass: the
shape was 1.1 mm long, one lip was 0.11 mm wide, the other was 0.20
mm wide; both lips appeared deflected with respect to the position
of the magnet-off bar, and the maximum gap between the upper and
lower lips was approximately of the order of magnitude of the width
of the lips. Stern and Gerlach made only visual observations through
a microscope, with no statistics on the distributions: they did not
obtain ``two spots'' as it is usually stated on textbooks, and
though the beam was clearly split in two distinguishable parts,
these were not disjoint. They accounted for the experiment as
exhibiting the property of ``space quantization in a magnetic
field.''

Let us give a simplified mathematical description of the experiment,
taking however into account not only the spin, but also the spatial
degrees of freedom of (the center of mass of) the silver
atoms~\cite{boo}. The silver atoms of a {Stern}-{Gerlach} experiment
can be treated as spin one-half elementary particles to a very high
degree of accuracy; let us assume that they are initially prepared
in an (arbitrary) spin state
\begin{equation} \label{eq:268}
\ket{c_{+}, c_{-}} = c_{+} \ket{+} + c_{-} \ket{-}, \qquad
\text{($|c_{+}|^{2} + |c_{-}|^{2} = 1$)},
\end{equation}
where~$\ket{+}$ and~$\ket{-}$ denote the usual normalized
eigenstates of~$S_{z}$. The wave function describing the center of
mass of the atom at the initial time~$t=0$ can be taken to be a
normalized Gaussian wave packet
\begin{equation} \label{eq:269}
G({\bf x}; {\bf x}_{0}, {\bf p}_{0}, \Delta_{\bf x}) =
  \braket{{\bf x} | {\bf x}_{0}, {\bf p}_{0}, \Delta_{\bf x}}
\end{equation}
centered around the position~${\bf x}_{0}$, traveling along the
$x$-axis towards the region where the magnet is located, with mean
value of momentum ${\bf p}_{0}$ and spatial spread~$\Delta_{\bf x}$,
extremely well localized with respect to the dimensions of the
region where the magnetic field is different from 0.

Let us model the interaction between the atom and the magnetic
field, which we treat as a static external field, by the usual
Hamiltonian operator $H_{\mathrm{SG}} = \vec{\mu}\cdot {\bf B}$ with
$\vec{\mu} = k \vec{\sigma}$, ${\bf B}$ the magnetic field,
$\vec{\sigma} = (\sigma_{x}, \sigma_{y}, \sigma_{z})$ the three
Pauli operators, and~$k$ a known constant whose value is irrelevant
for the following discussion.  Denoting by~$\hat{\bf z}$ the unit
vector along the $z$-axis, let us assume that the inhomogeneous
magnetic field is\footnote{This assumption is inconsistent with the
Maxwell equations, in particular with $\vec{\nabla} \times {\bf B} =
0$.  A more realistic assumption would be ${\bf B} = bx\,\hat{\bf x}
+ (B_{0} - b z)\,\hat{\bf z}$. A simple analysis shows, however,
that this second assumption gives rise to correcting terms which are
inessential for our present discussion.} ${\bf B} = (B_{0} - b
z)\,\hat{\bf z}$ inside the spatial region where the magnet is
located and with a negligible gradient outside it. Finally, to
simplify the matter as far as possible, let us adopt the ``impulsive
measurement'' assumption, which amounts to ignoring the free
evolution of the silver atom while it is interacting with the
magnetic field, and let us suppose that the time interval~$\tau$
between the emission of the silver atom from the oven and the moment
when it impinges on the glass plate is so small that the
spread~$\Delta_{\bf x}$ of its wave packet does not vary
appreciably.

With these premises, a simple calculation shows that the state of
the silver atom at the time~$t=\tau$, after it went through the
inhomogeneous magnetic field region and just before the detection by
the glass plate, is
\begin{equation} \label{eq:272}
  \begin{aligned}
    \ket{\psi_{\tau}} &
    = c_{+} \ket{{\bf x}_{+}, {\bf p}_{+}, \Delta_{\bf x}} \otimes
    \ket{+} + c_{-} \ket{{\bf x}_{-}, {\bf p}_{-}, \Delta_{\bf x}}
    \otimes \ket{-},
  \end{aligned}
\end{equation}
where the wave functions $G({\bf x}; {\bf x}_{\pm}, {\bf p}_{\pm},
\Delta_{\bf x}) = \braket{{\bf x} | {\bf x}_{\pm}, {\bf p}_{\pm},
\Delta_{\bf x}}$ are normalized Gaussian packets of mean momentum
${\bf p}_{\pm}$ and mean position ${\bf x}_{\pm}$ given
by\footnote{In deriving equation~\eqref{eq:273} the realistic
assumption $\tau k b \ll p_{0} \Rightarrow |{\bf p}_{\pm}| \simeq
|{\bf p}_{0}|$ (i.e., that the transverse variation of the momentum
is negligible with respect to its modulus) has also been taken into
account.}:
\begin{equation} \label{eq:273}
{\bf p}_{\pm} \; = \; {\bf p}_{0} \pm k b \tau\,\hat{\bf z},
\qquad\quad {\bf x}_{\pm} \; \simeq \; {\bf x}_{0} + \frac{{\bf
p}_{0}}{m}\,\tau \pm \frac{k b}{m}\,\tau^{2}\,\hat{\bf z}.
\end{equation}

The detection process by means of the glass plate can be modeled by
associating to the plate the operator $Z \equiv Z_{+} - Z_{-}$,
acting on the spatial degrees of freedom, where $Z_{+}$ and $Z_{-}$
are the two projection operators corresponding to the localization
in the upper ($z \geq 0$) or lower ($z \leq 0$) half parts of the
plate. Recalling that we choose a normalized vector to represent the
initial state, the corresponding probabilities according to standard
quantum mechanical rules are then easily seen to be
\begin{eqnarray} \label{eq:275}
{\mathbb P}_{(c_{+}, c_{-})}(\pm 1) & = &  \braket{\psi_{\tau} |
Z_{\pm}
\otimes I_{{\mathbb C}^{2}} | \psi_{\tau}} \nonumber \\
& = & |c_{+}|^{2} \braket{{\bf x}_{+}, {\bf p}_{+}, \Delta_{\bf x} |
Z_{\pm} | {\bf x}_{+}, {\bf p}_{+}, \Delta_{\bf x}} \; + \;
|c_{-}|^{2} \braket{{\bf x}_{-}, {\bf p}_{-}, \Delta_{\bf x}|
Z_{\pm} | {\bf x}_{-}, {\bf p}_{-}, \Delta_{\bf x}}
\end{eqnarray}

By looking at equation~\eqref{eq:275}, we immediately see that the
probability to observe an outcome in, say, the upper part of the
screen, is not $|c_{+}|^{2}$, as one can find in the textbook
descriptions of the experiment, because the two wave functions
$G({\bf x}; {\bf x}_{\pm}, {\bf p}_{\pm}, \Delta_{\bf x})$ are not
localized sharply in (respectively) the upper and lower parts of the
glass plate. Indeed, the tail of the ``+'' wave function $G({\bf x};
{\bf x}_{+}, {\bf p}_{+}, \Delta_{\bf x})$ which lies in the lower
part does not contribute at all to~${\mathbb P}_{(c_{+},
c_{-})}(+1)$, while the tail of the ``$-$'' wave function $G({\bf
x}; {\bf x}_{-}, {\bf p}_{-}, \Delta_{\bf x})$ laying in the upper
part does contribute (see Fig.~\ref{fig2}). Needless to say, this is
not a disfeature due to the use of a Gaussian wave-packet: the whole
analysis could be repeated practically unaltered for an arbitrary
initial wave packet, because even if one starts with a wave packet
with compact support, the free evolution would immediately and
unavoidably spread it all over space.
\begin{figure}
\begin{center}
{\includegraphics[scale=0.7]{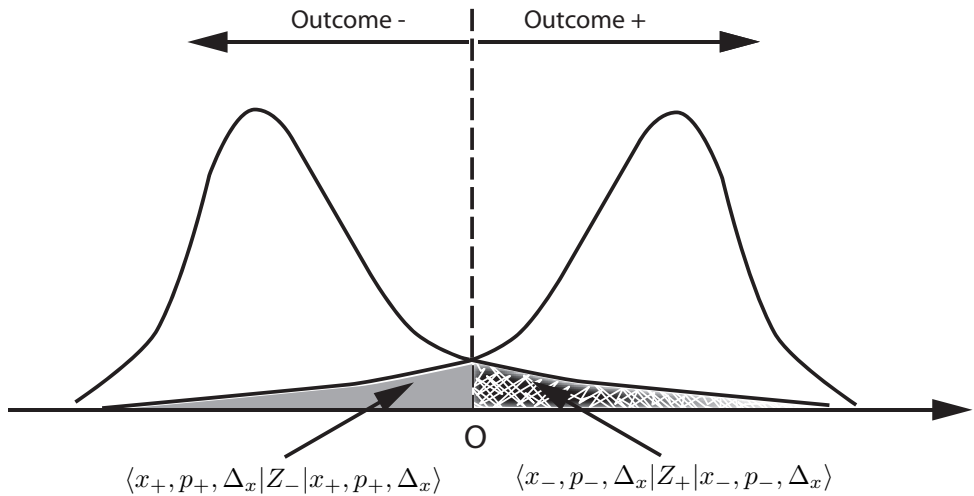}} \caption{Typical (schematic)
spatial distribution of the superposition of two Gaussian wave
functions in a Stern-Gerlach experiment. The wave function
corresponding to the outcome ``+'' has a tail spreading in the
region corresponding to the outcome ``$-$'', and vice-versa.}
\label{fig2}
\end{center}
\end{figure}

However, provided the initial wave packet is sufficiently well
localized both in position as well as in momentum (within the limits
allowed by the Heisenberg principle, of course), and the spatial
separation of the two final wave packets is much greater than their
spatial spreads, i.e. $2 k b \tau^{2}/m \gg \Delta_{\bf x}$ the
``crossed'' contributions of the tails, namely $\braket{{\bf x}_{+},
{\bf p}_{+}, \Delta_{\bf x} | Z_{-} | {\bf x}_{+}, {\bf p}_{+},
\Delta_{\bf x}}$ and $\braket{{\bf x}_{-}, {\bf p}_{-}, \Delta_{\bf
x} | Z_{+} | {\bf x}_{-}, {\bf p}_{-}, \Delta_{\bf x}}$, are small
compared with the uncrossed contributions, and the textbook
approximation ${\mathbb P}_{(c_{+}, c_{-})}(\pm 1) = |c_{\pm}|^{2}$
can be meaningfully recovered.

It is interesting to restate the previous argument in the following
way, by observing that equation~\eqref{eq:275} can be rewritten in
terms of the initial spin state~$\ket{c_{+}, c_{-}}$ as
\begin{equation} \label{eq:280}
{\mathbb P}_{(c_{+}, c_{-})}(\pm 1) = \Braket{c_{+}, c_{-} | O_{\pm}
| c_{+}, c_{-}}
\end{equation}
where the two effects $O_{\pm}$ are:
\begin{equation} \label{eq:281}
O_{\pm} \; \equiv \; \braket{{\bf x}_{+}, {\bf p}_{+}, \Delta_{\bf
x} | Z_{\pm} | {\bf x}_{+}, {\bf p}_{+}, \Delta_{\bf x}}
|+\rangle\langle+| \;+ \; \braket{{\bf x}_{-}, {\bf p}_{-},
\Delta_{\bf x}| Z_{\pm} |{\bf x}_{-}, {\bf p}_{-}, \Delta_{\bf x}}
|-\rangle\langle-|
\end{equation}
acting on the spin Hilbert space~${\mathbb C}^{2}$. We thus obtain a
result we are now used to: the probability distribution of the
outcomes has been expressed as the average value of the effects of a
POVM over the initial state of the microscopic quantum system.
However, since the wave packets are spread out in space, we see that
the effects~\eqref{eq:281} are {\it not} projection operators, i.e.,
$O_{\pm}^{2} \neq O_{\pm}$. Concluding, the observable associated to
the Stern-Gerlach experiment is not exactly represented by the
operator~$S_{z}$ (or, equivalently, by the PVM given by
$|+\rangle\langle+|$ and $|-\rangle\langle-|$): it is rather a
generalized observable described by the POVM formed by the two
effects~$O_{\pm}$. Thus, to be rigorous, also within standard
Quantum Mechanics the association of the operator~$S_{z}$ to a
Stern-Gerlach experiment is just an approximation, whose range of
validity has been made clear by the previous discussion: indeed,
when the crossed contributions of the tails of the two Gaussian wave
functions are negligible with respect to the uncrossed ones, we have
\begin{equation} \label{eq:282}
  O_{+} \; \simeq \; |+\rangle\langle+|
  \qquad\text{and}\qquad
  O_{-} \; \simeq \; |-\rangle\langle-|,
\end{equation}
so that the POVM $\{O_{+}, O_{-}\}$ associated to the experiment
approximately becomes the spectral PVM $\{|+\rangle\langle+|,
|-\rangle\langle-|\}$ of the spin operator~$S_{z}$.

\section{Observables as self-adjoint operators in classical mechanics: an
example from classical electromagnetic theory.} \label{sec:cla}

From the previous analysis one should have grasped that the operator
formalism for describing the outcome of experiments is not a
peculiar feature of Quantum Mechanics but, according to the Riesz
representation theorem, it can be applied whenever the states of
physical systems are represented by vectors of a linear vector space
and the outcome of an experiment depends ``quadratically'' on the
initial state of the system being measured. As such, there is no
reason why it should not be possible to apply such a formalism also
to classical systems: to illustrate this fact, we now give an
example taken from classical electromagnetic theory.

Let us confine our attention to a monochromatic plane wave of
frequency~$\omega$ traveling in the vacuum in some given
direction\footnote{According to the customary practice in classical
theories, in this subsection we will denote vectors with the
traditional bold notation~${\bf v}$ rather than the Dirac
notation~$\ket{v}$.}~$\hat{\bf n}$ (i.e., with wave vector ${\bf k}
= \omega \hat{\bf n}$).  Such a wave is completely characterized by
the electric field
\begin{equation} \label{eq:79}
{\bf E}(t, {\bf x}) = \text{Re}\, (\vec{\epsilon} \exp\left[
\ii\left( \omega t - {\bf k}\cdot {\bf x} \right) \right]);
\end{equation}
the complex wave amplitude~$\vec{\epsilon}$ expresses both the
intensity as well as the polarization state of the wave.
Accordingly, taking into account the linearity of Maxwell's
equations and the fact that the wave intensity\footnote{The wave
intensity we are referring here to is of course the mean value of
the Poynting vector of the wave over a time interval long enough
with respect to the wave period.}~$I$ can be written in terms
of~$\vec{\epsilon}$ by means of the canonical complex scalar
product,
\begin{equation}
I = \vec{\epsilon}^{*} \cdot \vec{\epsilon},
\end{equation}
we see that, if we limit our attention to the degrees of freedom
contained in the wave amplitude alone, ignoring the spatial ones, a
monochromatic plane wave qualifies as a physical system whose states
belong to the Hilbert space ${\mathbb C}^2$ and on which experiments
(typically detection of the intesity) represented by quadratic forms
can be performed.

One of the simplest examples of such a kind of experiment on a
monochromatic plane wave can be constructed by letting the wave
impinge on a linear Polaroid and by measuring the transmitted
intensity.  To further simplify the matter, let us restrict to the
case of a linearly polarized monochromatic plane wave, i.e., let us
assume $\vec{\epsilon} \in {\mathbb R}^{2}$.  Experience shows that
in such a situation the transmitted wave intensity~$I_{f}$ depends
on the angle between the directions of polarization of the wave and
the filter according to the Malus law:
\begin{equation}  \label{eq:81}
I_{f} \; = \; I_{i} \cos^{2}\theta,
\end{equation}
where $I_{i} = \vec{\epsilon} \cdot \vec{\epsilon}$ is of course the
incident wave intensity and, if $\hat{\vec{\eta}}$ is the
polarization direction of the filter, $\cos\theta = \vec{\epsilon}
\cdot \hat{\vec{\eta}} / \| \vec{\epsilon} \|$. Accordingly, Malus
law~\eqref{eq:81} can be re-written as,
\begin{equation}
I_{f} \; = \; (\hat{\vec{\eta}} \cdot \vec{\epsilon})^2,
\end{equation}
and we easily recognize that the outcome of the experiment (the
output intensity $I_{f}$) is a quadratic form of the initial state
of the wave (its polarization $\vec{\epsilon}$). Then, according to
the Riesz representation theorem, we can express $I_{f}$ in terms of
a bounded linear operator $O$ as follows (we now pass to the Dirac
notation to highlight the conclusion):
\begin{equation} \label{eq:85}
I_{f}(\hat{\vec{\eta}}) \; = \; \braket{\vec{\epsilon} |
O_{\hat{\vec{\eta}}} | \vec{\epsilon}} \qquad\quad
O_{\hat{\vec{\eta}}} \; = \; |\hat{\vec{\eta}} \rangle\langle
\hat{\vec{\eta}}|.
\end{equation}
Of course, in the present example, the average value of the operator
$O_{\hat{\vec{\eta}}}$ over the initial state of the micro-system
does not represent the average of the possible outcomes weighted
with their probabilities, but it is simply the (deterministic) value
of the output intensity.

Let us note that this example also shows that
\emph{non-commutativity,} which is usually considered as a
characteristic trait of quantum mechanics, can and does arise in
classical and deterministic contexts as well.  Indeed, by taking
into account the fact that the transmitted wave is polarized along
the direction of the filter:
\begin{equation} \label{eq:86}
\ket{\vec{\epsilon}_{\text{out}}} \; = \; \ket{\hat{\vec{\eta}}}
\braket{\hat{\vec{\eta}} | \vec{\epsilon}_{\text{in}}},
\end{equation}
and by considering two successive arbitrarily oriented filters, we
see that the transmitted intensity
\begin{equation} \label{eq:87}
I_{f} = I_{i} \cos^{2}\theta_{1} \cos^{2}(\theta_{2} - \theta_{1})
\end{equation}
of the wave after the two filters {\it depends} in general on their
respective order. The mathematical counterpart of this property is
that the two operator $O_{\hat{\vec{\eta}}_{1}}$ and
$O_{\hat{\vec{\eta}}_{2}}$ associated to the two experiments do {\it
not} commute, a part from the very special orientations which
correspond to $\ket{\hat{\vec{\eta}}_{1}} = \pm
\ket{\hat{\vec{\eta}}_{2}}$ or $\braket{\hat{\vec{\eta}}_{1} |
\hat{\vec{\eta}}_{2}} = 0$.

\section{Conclusions}
\label{sec:conclusions}

In this paper we have shown how the Hilbert-space operator
formalism---i.e., the use of self-adjoint operators and, more
generally, of POVM to describe experiments on quantum systems---can
be recovered within the context of DRMs: it can be derived from the
dynamical laws of DRMs satisfying the very general assumptions that
we have analyzed in
section~\ref{sec:hsop-drm:outcomes_microstates-link} as a compact
way to express the statistical properties of the outcomes of
\emph{measurement-like} experiments. It is worthwhile stressing once
again that within the context of DRMs the operator formalism has no
special ontological meaning: as this paper shows in detail, it is
merely a convenient tool that can be used to describe certain
experiments that we are accustomed to think of as ``measurements.''

This and the previous analysis prove that DRMs represent a
well-grounded theory, whose ontology is clearly specified and
consistent with our macroscopic perceptions, deprived from the
typical paradoxes of quantum mechanics which are connected to the
special roles that the theory attributes to ``measurement
processes'' and ``observables''.

We consider the results of this paper as an important step which
completes, at the non-relativistic level, the general line of though
and the world view which has inspired the dynamical reduction
program, both at the formal and interpretational level.

\begin{acknowledgments}
We are indebted with D. D\"urr and R. Tumulka for many stimulating
comments. DGMS gratefully acknowledges support by the Department of
Theoretical Physics of the University of Trieste. The work of AB has
been supported partly by the EU grants MEIF\;CT\;2003--500543 and
ERG\;044941-STOCH-EQ, and partly by DFG (Germany).
\end{acknowledgments}

\end{document}